# Image Processing with Dipole-Coupled Nanomagnets: Noise Suppression and Edge Enhancement Detection

Md Ahsanul Abeed, Ayan K. Biswas, Md Mamun Al-Rashid, Jayasimha Atulasimha and Supriyo Bandyopadhyay

*Abstract*— Hardware based image processing offers speed and convenience not found in software-centric approaches. Here, we show theoretically that a two-dimensional periodic array of dipole-coupled elliptical nanomagnets, delineated on a piezoelectric substrate, can act as a dynamical system for specific image processing functions. Each nanomagnet has two stable magnetization states that encode pixel color (black or white). An image containing black and white pixels is first converted to voltage states and then mapped into the magnetization states of a nanomagnet array with magneto-tunneling junctions (MTJs). The same MTJs are employed to read out the processed pixel colors later. Dipole interaction between the nanomagnets implements specific image processing tasks such as noise reduction and edge enhancement detection. These functions are triggered by applying a global strain to the nanomagnets with a voltage dropped across the piezoelectric substrate. An image containing an arbitrary number of black and white pixels can be processed in few nanoseconds with very low energy cost.

## I. INTRODUCTION

HARDWARE based image processing can be extremely fast and yet consumes little energy. Here, we propose and analyze a magnetic system for this purpose and show that it can carry out specific image processing tasks in a few nanoseconds while consuming miniscule amount of energy.

The choice of nanomagnets for image processing is motivated by the desire for energy efficiency. Traditional hardware platforms for image processing employ field-programmable gate arrays [1], diodes [2], nanowires [3] or quantum dots [4]. The image processing function is implemented via charge exchange between these elements, which always results in current flow and associated $I^2R$ power loss. In contrast, no current has to pass through nanomagnets in an image processing function, which eliminates the $I^2R$ loss in the device. Furthermore, the interaction between nanomagnets is "wireless" (no physical wires) and is mediated by dipole interaction. Hence there is no energy loss in interconnects either, which further improves the energy efficiency. These considerations make nanomagnetic systems a very attractive choice for image processing.

## II. THEORY

In this paper, we will consider processing of black and white images only, i.e. each pixel is either black or white. We will encode a pixel's color – black or white – in a stable magnetization orientation of a nanomagnet. This will require the nanomagnet to have only two stable magnetization orientations. The simplest way to achieve that is to choose nanomagnets that are shaped like an elliptical disk. A static fixed magnetic field is directed along the minor axis of the ellipse as shown in Fig. 1(a). That will ensure that the nanomagnet's magnetization will have only two stable orientations in the ellipse's plane as shown in Fig. 1 [5]. The angle $\varphi$ between them is a function of the strength of the magnetic field and the eccentricity of the ellipse. It can be adjusted to $\sim 90^0$ by tuning the magnetic field strength [5][6].

Next consider a two-dimensional array of nanomagnets in a global magnetic field directed along the ellipses' minor axes as shown in Fig. 1(b). Each nanomagnet will now experience a dipole field due to dipole interaction with its neighbors. This field depends, among other things, on the nanomagnet's own magnetization state, as well as the states of its neighbors. The effect of the dipole field on any nanomagnet is to make one of the two stable magnetization orientations preferred over the other. Which one is preferred will depend on the magnetization states of the neighbors. Therefore, the preferred color (black or white) of the pixel encoded in any nanomagnet will be dictated by the colors of the surrounding pixels. *This dependency results in specific image processing functions.*

### A. Triggering the image processing action

The image processing function requires a nanomagnet to transition to its preferred magnetiztion orientation. However, this does not happen automatically. The potential energy profile of an *isolated* nanomagnet has two degenerate minima corresponding to the two stable magnetization orientations. When a nanomagnet is placed in the vicinity of others, the

This work is supported in part by the Commonwealth of Virginia, Center for Innovative Technology (CIT) via the Commonwealth Research Commercialization Fund (CRCF) Matching Funds Program MF-15-006-MS.

M.A.A, A.K.B. and S.B. are with the Department of Electrical and Computer Engineering, Virginia Commonwealth University, Richmond, VA 23284 USA (e-mail: abeedma@vcu.edu, biswasak@vcu.edu, sbandy@vcu.edu).

M.M.A and J.A. are with both the Department of Electrical and Computer Engineering and the Department of Mechanical and Nuclear Engineering, Virginia Commonwealth University, Richmond, VA 23284 USA (e-mail: alrashidmm@vcu.edu, jatulasimha@vcu.edu)

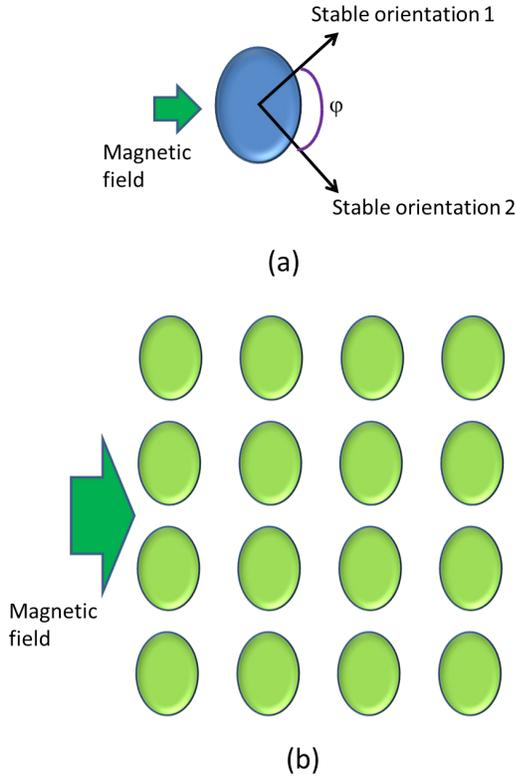

Fig. 1. (a) An isolated elliptical nanomagnet, placed in a magnetic field directed along the minor axis of the ellipse, has two stable magnetization orientations that are equally preferred. The angle $\varphi$ between them can be adjusted by varying the strength of the magnetic field. (b) A two-dimensional array of elliptical nanomagnets in a global magnetic field. Dipole interaction makes one stable orientation preferred over the other and which is the preferred orientation is determined by the orientations of the neighbors.

dipole interaction makes one of the two minima lower in energy than the other, thereby making one stable orientation (corresponding to the lower energy minimum) preferred over the other. For the nanomagnet's magnetization to reorient in the more preferred direction, it must be able to transition to the lower energy minimum. That may not happen if there are energy barrier(s) separating the higher (local) energy minimum from the lower (global) energy minimum. In that case, an external agent must be employed to erode the energy barrier(s) temporarily and allow the system to migrate to the lower energy state, thus completing the image processing function. The external agent acts as a "clock" to trigger the image processing activity.

*B. Strain as the trigger*

Many external agents can act as the clock. However, the most energy-efficient clocking agent is mechanical strain [7][8] which lowers the potential barrier between the stable magnetization orientations in a magnetostrictive nanomagnet and switches the magnetization from the non-preferred orientation to the preferred one in the presence of inter-magnet dipole interaction. Strain is generated in the following way. The nanomagnets are delineated on a piezoelectric substrate and a voltage is applied across the latter with electrodes delineated on the surface of the substrate. By placing the electrodes in an appropriate fashion, one can generate biaxial strain (tensile along the major axis of the elliptical nanomagnet and compressive along the minor axis, or vice versa, depending on the polarity of the voltage). This strain is partially or wholly transferred to the nanomagnets and reorients each one's magnetization to the preferred direction [9][10][11], thus completing the image processing action.

*C. Reading and writing of pixels into nanomagnets*

For such a paradigm to work, we must first be able to write pixel information into the magnetization states and then be able to read them. For the purpose of "writing", we will fabricate a skewed magneto-tunneling junction (MTJ) on top of the elliptical nanomagnet as shown in Fig. 2. The nanomagnet will act as the soft layer of the MTJ and the hard layer of the MTJ (implemented with a synthetic anti-ferromagnet) will be permanently magnetized along one of the two stable magnetization directions of the soft layer.

A negative potential applied between the hard layer and soft layer (negative polarity of the voltage source connected to the hard layer) will inject electrons from the hard into the soft layer and orient that latter's magnetization along that of the hard layer by spin-transfer torque. This will, say, write a black pixel state into the soft layer. Reversing the polarity of the potential will extract electrons from the soft layer into the hard layer and switch the magnetization of the soft layer to the other stable state, thus writing the white pixel. The actual pixel color can be converted to a voltage with a photodetector that generates a voltage proportional to the brightness (white pixels high voltage and black pixels low voltage). A level shifter then transduces a white pixel state to a positive voltage and a black pixel state to a negative voltage, resulting in direct conversion of pixel color to magnetization state.

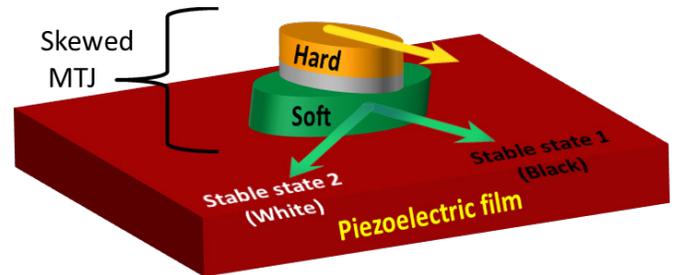

Fig. 2. A skewed magneto-tunneling junction whose soft layer has two stable magnetization orientations that correspond to black and white pixels. A magnetic field directed along the minor axis of the soft layer is present and not shown. The hard layer is permanently magnetized along one of the stable orientations. A black or white pixel state can be stored or written by applying a negative or positive voltage between the hard and soft layers with contacts not shown. The same contacts can be used to read the resistance of the MTJ and deduce the stored pixel color.

To "read" a stored pixel state, we will measure the corresponding MTJ's resistance. A low resistance state will imply that the magnetizations of the hard and soft layer are parallel and hence the stored pixel is black, while a high resistance state will imply that the two magnetization states are approximately perpendicular and the stored pixel is white.

Having established the read/write scheme, we will now proceed to discuss the image processing functionality. We first discuss the nanomagnet properties in Section III and present the image processing results in Section IV.

## III. NANOMAGNET PROPERTIES

The magnetostrictive nanomagnets (soft layers of the MTJ) in this study are assumed to be made of Terfenol-D which is among the most magnetostrictive materials known. The dimensions of the elliptical nanomagnets are 100 nm (major axis), 60 nm (minor axis) and 16 nm (thickness). Table I lists the properties of the nanomagnets.

TABLE I
NANOMAGNET PARAMETERS [12][13][14][15]

| Symbol | Quantity | Value |
| --- | --- | --- |
| $M_s$ | Saturation Magnetization | $8 \times 10^5$ A/m |
| $A$ | Exchange parameter | $9 \times 10^{-12}$ |
| $\lambda_s$ | Magnetostriction coefficient | 600 ppm |
| $B$ | Global magnetic field | 82 mT |

We will designate the magnetization orientation of a nanomagnet (or, equivalently, the pixel color encoded in that nanomagnet) by the angle $\theta$ subtended with the major axis of the ellipse. If we assume that the $z$-axis is along the major axis of the ellipse and the $y$-axis is along the minor axis, then in spherical coordinates, $\theta$ is the polar angle and we will call the azimuthal angle $\phi$. For the nanomagnet dimensions and the magnetic field strength given in Table I, the two stable magnetization orientations are computed to be ($\theta = 48.6^o$, $\phi = 90^o$) and ($\theta = 131.4^o$, $\phi = 90^o$) following the method of [6]. The angular separation between them ($\varphi$) is $82.8^o$. These two orientations will encode "black" and "white", respectively.

In Fig. 3, we show the two stable magnetization orientations and the potential energy profile of an isolated nanomagnet (as a function of $\theta$, assuming $\phi = 90^o$) in the absence of any stress or dipolar interaction from neighbors. As expected, there are two degenerate minima at $\theta = 48.6^o$ and $131.4^o$, indicating that these are the stable orientations. The potential barrier between them is 42 $kT$ ($T$ = 300 K). This figure is calculated from the expression for the potential energy as a function of $\theta$ given in [8].

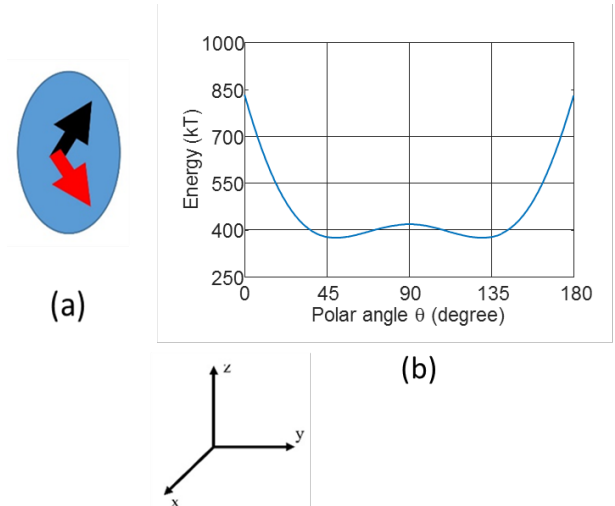

Fig. 3. (a) The two stable magnetization orientations. (b) The potential energy profile in the magnet's plane ($\phi = 90^0$) as a function of the polar angle $\theta$ of the magnetization. The barrier separating the two degenerate minima is 42 $kT$ ($T$ = 300 K).

In the presence of stress, dipolar interaction and thermal noise, the spins in each magnetostrictive nanomagnet experience six different magnetic fields: a demagnetization field due to the nanomagnet's (elliptical) shape anisotropy, a dipolar field arising from dipole interaction due to the nanomagnet's neighbors, the global magnetic field that defines the two stable states of a nanomagnet, a field due to any stress generated in the magnetostrictive nanomagnet, a field due to exchange interaction between spins within a nanomagnet, and a random field due to thermal noise. Since there is no current flow through the nanomagnets, there is no Shot noise or flicker noise. We will ignore any field due to magneto-crystalline anisotropy since the nanomagnets are assumed to be amorphous. These six fields determine the spin texture (orientations of spins) within each nanomagnet and hence its magnetization orientation.

## IV. IMAGE PROCESSING

### A. Correcting single pixel error

Recall that the color "black" is encoded in the $\theta = 48.6^o$ state and the color "white" in the $\theta = 131.4^o$ state. Now, consider a black segment of an image consisting of 7×7 pixels where every, except one, pixel is black and the errant pixel is white. The white pixel represents noise or defect in the black segment of the image. The pixels are mapped into a 7×7 array of nanomagnets shown in Fig. 4(a) [using the writing scheme described earlier] where the vertical separation between the centers of two nearest neighbor nanomagnets is 150 nm and the horizontal separation is 285 nm. Because of dipole interaction from the neighbors, the potential energy profile of the errant nanomagnet looks like that in Fig. 4(b), where the $\theta = 48.6^o$ state is actually slightly *lower* in energy than the $\theta = 131.4^o$ state and there is an energy barrier of 4.6 $kT$ at $\theta = 90^o$ separating the two minima. The dipole interaction has lifted the degeneracy of the two stable states and made the "black" state *preferred* over the "white". Dipole interaction actually also changes the two stable orientations slightly from $\theta = 48.6^o$, $131.4^o$ (which are the stable orientations of an isolated nanomagnet), but since this is a small effect and of no import, we will continue to label the stable orientations as $\theta \approx 48.6^o$ and $\theta \approx 131.4^o$ states.

We would expect that the errant nanomagnet would prefer to migrate to the lower energy state and the corrupted white pixel would spontaneously turn black, thereby auto-correcting and eliminating noise by virtue of the dipole interaction. This actually does happen in this case since the energy barrier separating the two minima is a mere 4.6 $kT$ and thermal fluctuations can transcend this barrier. Therefore, the corrupted pixel auto-corrects. However, this would not have happened if the barrier were much higher than 4.6 $kT$. In that case, the nanomagnet would have remained stuck in the higher energy metastable state (white pixel) and could not decay to the ground state (black pixel), thereby preventing error correction. To remedy this situation, a compressive global stress can be applied along the major axis of the ellipses as shown in Fig. 4(a). The stress adds an extra term to the potential energy of every nanomagnet: $E_{stress} = -\frac{3}{2}\lambda_s \sigma \Omega \cos^2 \theta$, where $\lambda_s$ is the magnetostriction coefficient of the magnet (it is a positive quantity for Terfenol-D), $\sigma$ is the stress (positive for tension and negative for compression) and $\theta$ is the angle between the stress axis and magnetization, which also happens to be the polar angle of the magnetization vector. This energy term is minimum when $\theta = 90°$, which means that compressive stress would remove the energy barrier at $\theta = 90°$ and allow the corrupted pixel to auto-correct.

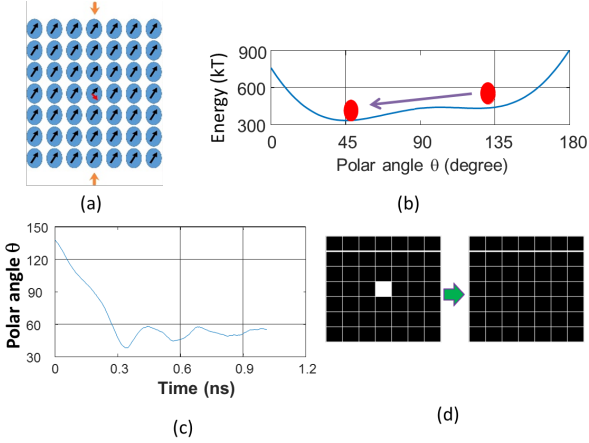

Fig. 4. (a) A $7\times 7$ array of nanomagnets storing a $7\times 7$ array of pixels (not to scale). All pixels are black except the one in the center which has turned white due to noise. (b) The potential energy profile of the errant nanomagnet in the presence of dipole interaction with the neighbors. The degeneracy between the two minima has been lifted by dipole interactions. (c) Thermal fluctuations are able to take the errant nanomagnet to the global minimum in ~0.4 nanoseconds without the application of any stress. (d) The corrupted white pixel spontaneously turns black, thereby automatically removing the noise.

### B. Two neighboring pixels in error

In Fig. 5, we show the spin textures of nanomagnets in a $7\times 7$ array corresponding to the black segment ($7\times 7$ pixels) of an image where *two* vertically neighboring pixels in the fourth column have been corrupted and become white. In this case, thermal fluctuations alone cannot auto-correct the errant pixels since the barrier separating the two energy minima in the potential profiles of these two nanomagnets is too high for thermal fluctuations to overcome. However, application of a global compressive stress of magnitude 10.5 MPa along the major axes of the ellipses suppresses the energy barriers and rotates the magnetization of every nanomagnet. Subsequent removal of stress changes the magnetization states of the two errant nanomagnets *only*, so the colors of the two corrupted pixels are restored to approximately the right color while the colors of the uncorrupted pixels are left practically intact.

The micromagnetic simulations to obtain the results in Fig. 5 were carried out with the MuMax3 package [16]. The cell size used in the simulation was progressively decreased until the results became independent of cell size. This cell size was 3 nm $\times$ 3 nm $\times$ 2 nm.

Note two important features in Fig. 5. First, not all the spins in a nanomagnet representing a black pixel are pointing in exactly the $\theta = 48.6°$ direction. This happens because exchange interaction between the spins will always slightly reorient some spins to minimize the exchange energy. The same will be true for the spin texture in a nanomagnet representing a white pixel, i.e. not every spin will be pointing in exactly the $\theta = 131.4°$ direction. The spin texture will also fluctuate slightly with time because of thermal noise, but this does not impact the image processing function in any way.

Second, note that the black pixels in the final processed image are not identical to those in the initial image, but are very close to them. Therefore, if we digitize the read-out voltages, then the close-to-black pixels will be interpreted as black and the de-noising would be perfect. Even otherwise, the large brightness gradient at the center of the image has been eliminated and the overall noise reduced.

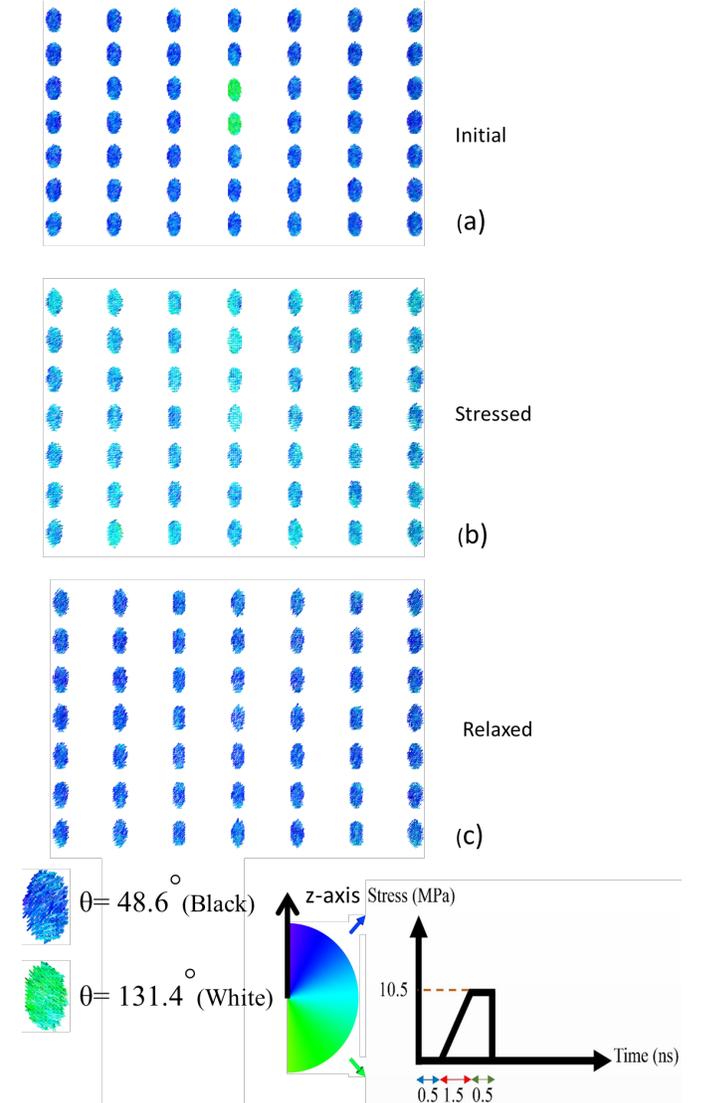

Fig. 5. (a) Spin texture of a $7\times 7$ pixel array representing the black segment of an image where two consecutive pixels in the fourth column are of the wrong color. (b) Spin textures of the array after application of 10.5 MPa of compressive stress along the major axes of the ellipses. (c) Spin textures after withdrawal of stress showing that the errant pixels are no longer white. All pixels have turned approximately black. In the bottom panel, we show (from left to right) spin texture of nanomagnets encoding black and white pixels, the color coding scheme ($\theta = 0°$ is blue and $\theta = 180°$ is green; intermediate colors represent $\theta$ values between $0°$ and $180°$), and the applied stress profile in time.

Note that since the applied stress is *global*, it affects every nanomagnet in the array. Therefore, every pixel changes color - the corrupted ones change significantly from white to near-black, while the uncorrupted ones change slightly from black to near-black. In Fig. 6, we show the time evolutions of the magnetization orientations (polar angle $\theta$) of the two nanomagnets encoding the corrupted pixels, as well as that of a nanomagnet encoding an uncorrupted pixel. The applied stress profile is shown at the bottom of Fig. 5. Note that the uncorrupted nanomagnet returns close to its original state of $\theta = 48.6°$ after completion of the stress cycle while a corrupted one changes significantly from $\theta \approx 131.4°$ to $\theta \approx 48.6°$.

The switching trajectories shown in Fig. 6 will vary from one run to another of the MuMax3 simulation because the field acting on the nanomagnets due to thermal noise is random in time. That is why we have plotted six different traces, each corresponding to a different simulation run. Note that although the trajectories vary somewhat because of thermal noise, each trajectory inevitably results in successful evolution of the errant nanomagnet from $\theta \approx 131.4°$ to $\theta \approx 48.6°$.

We have repeated the simulations for the converse situation when the two corrupted pixels are black and the surrounding ones are white (this would correspond to the white segment of an image). The black pixels turn nearly white after stress application and the white pixels remain nearly white, showing that error correction is independent of pixel color.

*C. Entire rows in error*

In Fig. 7(a), we show that if an entire row is corrupted, then the application of global stress along the major axes of the elliptical nanomagnets can correct the corrupted row. In Fig. 7(b), we show that the same is true if two consecutive rows are corrupted. In Figs. 7(c) and 7(d), we show that we can even correct two non-consecutive rows, regardless of whether the corrupted pixels are black or white.

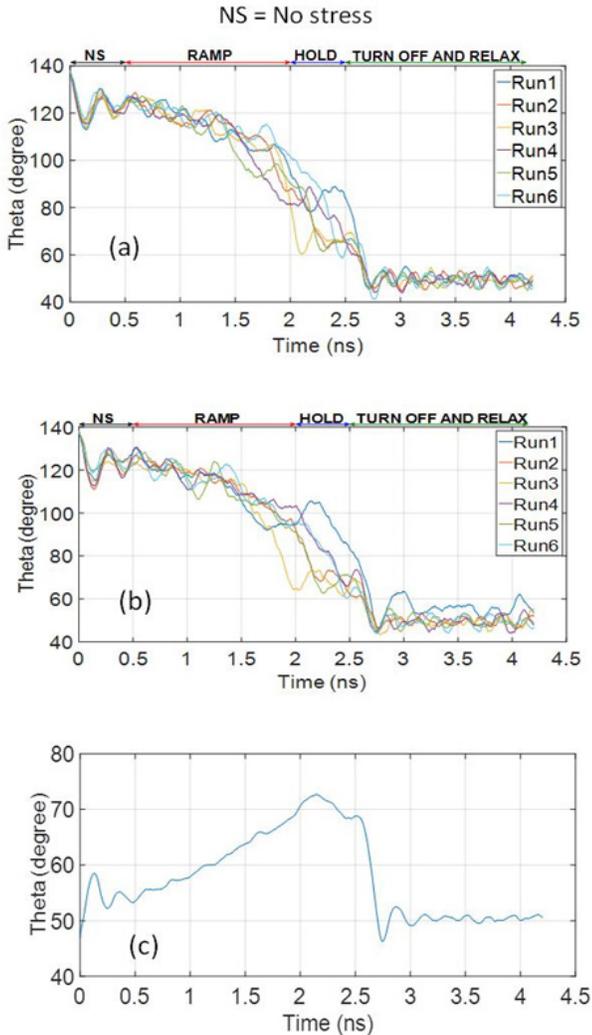

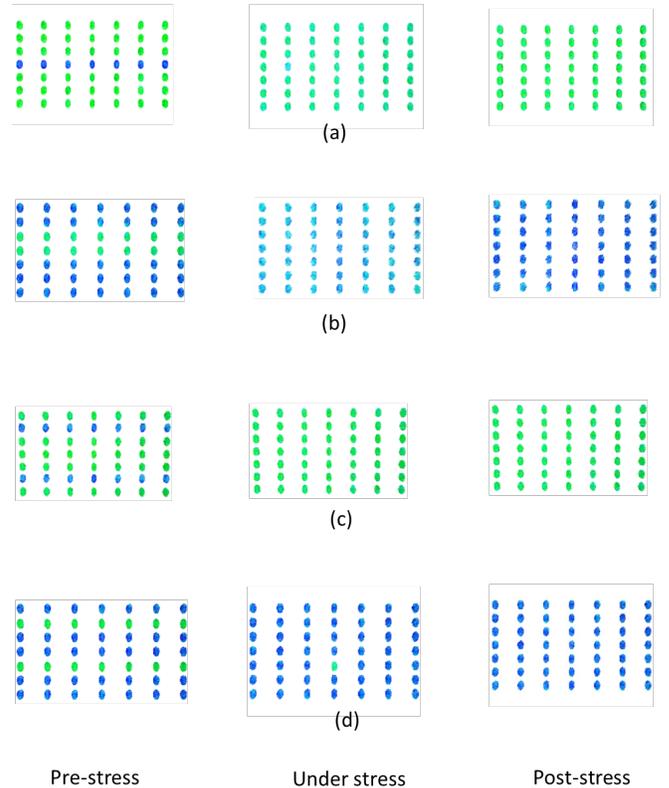

Fig. 6. Temporal evolution of the magnetization orientation of the (a) bottom and (b) top errant nanomagnet in Fig. 5 from $\theta \approx 131.4°$ to $\theta \approx 48.6°$. Since the magnetic field due to thermal noise is random, we ran six different simulations for each of the two nanomagnets. The six switching trajectories vary slightly but all result in transitioning to the correct pixel color corresponding to $\theta = 48.6°$. (c) The temporal evolution of the magnetization of the nanomagnet in the top right corner (row 1, column 7) in Fig. 5. This nanomagnet corresponds to an uncorrupted pixel, but because it too is subjected to global stress and thermal noise, its magnetization changes with time. However, it returns to its original state after a brief sojourn showing that the stress does not corrupt an uncorrupted pixel. The time to settle (image processing time) is ~3 nanoseconds.

Fig. 7. (a) Spin textures of an array of $7 \times 7$ nanomagnets representing white segment of an image where the entire fourth row has been corrupted and turned black. Application and subsequent withdrawal of global stress along the major axes corrects the corrupted row. (b) An array of black pixels in the black segment of an image where two consecutive rows have been corrupted and turned white. Stress application and withdrawal corrects both corrupted rows. (c) An array of white pixels where two non-consecutive rows have been corrupted and turned black. A stress cycle corrects both rows. (d) An array of black pixels where two non-consecutive rows have been corrupted and turned white. Again, a stress cycle corrects both rows.

## D. Random pixels in error

In Fig. 8, we show that even if random pixels are corrupted, as long as the corrupted pixels are a minority, application and withdrawal of global stress can correct the corrupted pixels.

## E. Edge enhancement detection

In Fig. 9, we show that if an image has two segments such that in one segment black pixels dominate and in the other white pixels are more abundant, then application of global stress turns the first segment all-black and the second segment all-white. This enhances the contrast between the two segments and sharpens the edge between them.

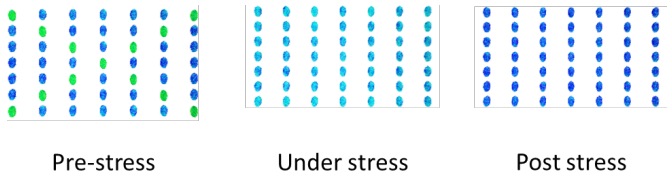

Fig. 8. A 7×7 array corresponding to the black segment of an image with randomly corrupted pixels (black pixels turned white). Application and subsequent removal of global stress applied along the major axes of the elliptical nanomagnets corrects the corrupted pixels and turns them all black.

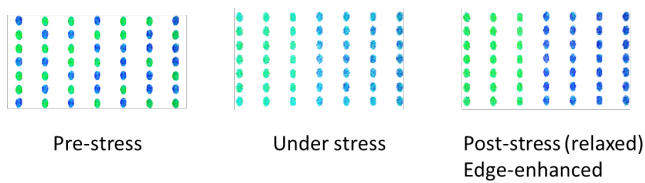

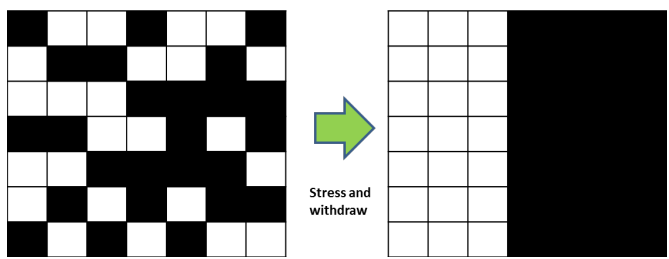

Fig. 9. Edge enhancement detection. The left segment (three columns) has majority white pixels and the right segment (four columns) has majority black segments. After application and removal of global stress applied along the major axes of the elliptical nanomagnets, the white-dominant segment turns all-white and the black-dominant segment turns all-black. This enhances the edge and contrast between the two segments.

## F. Exceptions and failures

There are some cases where the approach fails, an example

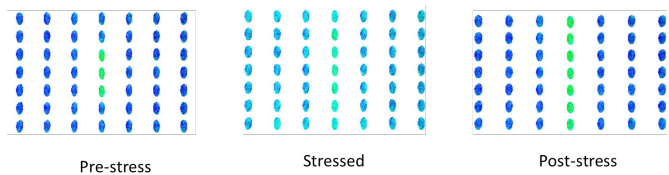

Fig. 10. Failure to correct pixels. The left panel shows a 7×7 array where three consecutive members of a column have been corrupted. Application and subsequent removal of global stress along the major axes of the ellipses corrupts the entire column.

of which is shown in Fig. 10.

What causes the failure in this case is the nature of dipole coupling. If the line joining the centers of two nanomagnets is collinear with their major axis, then dipole coupling will tend to make their magnetizations mutually parallel (ferromagnetic ordering), whereas if that line is collinear with the minor axis, then dipole coupling will make their magnetizations anti-parallel (ant-ferromagnetic ordering). The ferromagnetic coupling is stronger than the anti-ferromagnetic coupling [17]. Here, three consecutive errant nanomagnets are ferromagnetically coupled in the fourth column. There are two ferromagnetically coupled nanomagnets above and two below in that column which are in the correct state. The ferromagnetic coupling of the three in the incorrect states overwhelms that of the two above and two below in the correct state and ultimately corrupts the entire column to make it ferromagnetically ordered, i.e. every nanomagnet in the column is magnetized in the same direction, albeit the wrong direction. Thus, the approach fails in the rare case when consecutive pixels in a column are corrupted and the uncorrupted pixels above or below are fewer in number than the corrupted pixels.

## V. CONCLUSION

In conclusion, we have demonstrated a paradigm for removing image noise and edge enhancement detection using an array of dipole-coupled magnetostrictive nanomagnets whose magnetizations are sensitive to stress. There is significant interest in all-hardware based image processing [1][2][3][4][17][18] because of the speed and convenience. The present system serves this purpose for specific applications. The time taken to complete the image processing tasks considered here is a few nanoseconds and will be relatively independent of image size (number of pixels) since corrupted pixels are corrected simultaneously and not sequentially, i.e. the processing is parallel and not serial. Equally important, the energy dissipated to complete these tasks will be very small since very little energy is dissipated to switch the magnetization of magnetostrictive nanomagnets, delineated on a piezoelectric substrate, with electrically generated strain [19][20]. This results in an image processing paradigm with exceptionally low energy-delay product.